\documentclass[epj]{svjour}
\usepackage{latexsym}
\usepackage{graphics}
\usepackage{graphicx}
\begin{document}
\bibliographystyle{epjunsrt}

\title{Environment influence on the IR fluorescence of Xe$_{2}^{*}$ molecules in
electron beam excited Ar--Xe mixture at high density}

\author{A.F.Borghesani\inst{1}\thanks{\emph{E-mail address:} {\tt borghesani@padova.infm.it}} \and G.Carugno\inst{2} \and D.Iannuzzi\inst{3}\and I.Mogentale\inst{4}
}
\institute{Istituto Nazionale per la Fisica della Materia,
Dept. of Physics, University of Padua, via
F.Marzolo 8, I-35131 Padua, Italy \and Istituto Nazionale di
Fisica Nucleare, via F.Marzolo 8, I-35131 Padua, Italy \and
Harvard University,
Division of Engineering and Applied Sciences, 9 Oxford St., Cambridge, MA 02138, U.S.A. \and Istituto Nazionale di Fisica Nucleare,
Dept. of Physics, University of Padua, via
F.Marzolo 8, I-35131 Padua, Italy}

\date{Received: \today / Revised version: date}

\abstract{We report new measurements of the near infrared (NIR) Xe$_2^*$ excimer fluorescence in an electron--beam--excited Ar (90\%)--Xe (10 \%) mixture at room temperature. Previous measurements up to a density $N\approx 2\times 10^{26}$ m$^{-3}$ discovered a broad excimer fluorescence band at $\approx 7800$ cm$^{-1},$ whose center is red--shifted by increasing $N$ (A. F. Borghesani, G. Bressi, G. Carugno, E. Conti, and D. Iannuzzi, {\em J. Chem. Phys.}, {\bf 115}, 6042 (2001) \cite{Bor}).  The shift has been explained by assuming that the energy of the optical active electron in the molecule is shifted by the density--dependent Fermi shift and by accounting for the solvation effect due to the environment. We have extended the density range up to $N\approx 6\times 10^{26}$ m$^{-3}, $ confirming the previous measurements and extending the validity of the interpretative model. A detailed analysis of the width of the fluorescence band gives a value of 2.85 nm  for the size of the investigated excimer state. Such a large value lends credence to the validity of the proposed explanation of the experimental findings. 
} 
\titlerunning{Enviroment influence on the IR fluorescence of Xe$_{2}^{*}$ molecules}
 \maketitle
 
 \section{Introduction}\label{intro}
Irradiation of a noble gas sample with ionizing particles \cite{kogo,nowak1,galy,kannari}, with lasers \cite{raymond84,museur94}, or with synchroton light \cite{dutuit78} may produce, in addition to excited and ionized atomic species, molecular excited states, such as Xe$_2,$ for instance, which are called excimers. Upon decay, excimers release a large fraction of the absorbed energy in the vacuum ultraviolet (VUV) range that is exploited in many applications. For this reason noble gas excimers play a significant role in many areas of physics.

In the laser physics area, ultraviolet-- and vacuum ultraviolet lasers exploit excimers as light--amplifying media or as intermediate products in the kinetic chain leading to the population of the lasing level \cite{galy,exc}.

Excimer VUV fluorescence of noble gases is also used in scintillation particle detectors \cite{knoll}. Fluorescence is the final step of the degradation of the energy released in the detector medium by an ionizing particle. The transition from the first excited excimer level to its dissociative ground state produces the VUV radiation \cite{koe} that signals the passage of the energetic particle. 

Among rare gases, Xe is particularly studied for several reasons. For par\-ti\-cle--de\-tec\-tion purposes, in fact, it combines a large atomic weight with a high scintillation yield and its VUV fluorescence occurs at the longest wavelength, thus making it more technologically amenable. Mo\-re\-o\-ver, from a fundamental point of view, its excimers are more complex than those of other noble gases because of the larger spin--orbit coupling and overlap of atomic configurations.

The physical processes leading to excimer fluorescence in rare gases in low density limit are very well known \cite{nowak1,galy,sauer85,audouard88,nowak2,keto97,mulliken70,moutard87,ulrich87,salamero81,millet78}.  
The potential curve of diatomic rare gas molecules in the ground state is repulsive except for a weak van der Waals attraction.
An excited electronic state of the molecule can be obtained by promoting an electron to a Rydberg--like molecular orbital, whose nature may be either bonding or repulsive. Owing to the dissociative nature of the ground state, transition to it from a higher lying excimer level gives origin to fluorescence bands, whose width is related to the steepness of the potential of the molecular ground state. 

In the case of Xe, two such VUV bands, named {\em first--} and {\em second continuum}, have been observed \cite{museur94}. The first excited excimer levels Xe$_2^*$ $(1_u, 0^-_u)$ and Xe$_{2}^{*}$ $(0_{u}^{+})$ (in the
Hund's case {\em c}) or $^{1}\Sigma_{u}^{+}$ and $^{3}\Sigma_{u}^{+}$
(in the notation, in which spin-orbit coupling is neglected)
are produced in the collision of a neutral atom in its ground state Xe $(^1S_0)$ with atoms excited in the states 6s[3/2]$^o_2$ and 6s[3/2]$^o_1$ (in Racah notation), according to the reaction scheme \cite{moutard88}
\begin{eqnarray}
\mbox{Xe}\, (6\mbox{s}[3/2]^o_2) + 2 \mbox{Xe}\, (^1S_0)& \rightarrow &\mbox{Xe}_2^*\,
(1_u ,0^-_u) +  \mbox{Xe}\, (^1S_0) \label{eq:exc1}\\
\mbox{Xe}\, (6\mbox{s}[3/2]^o_1) + 2 \mbox{Xe}\, (^1S_0)& \rightarrow & \mbox{Xe}_2^*\,
(0_u^+) +  \mbox{Xe}\, (^1S_0) \label{eq:exc2}
\end{eqnarray}
The symbol $^*$ indicates an excited state.

Subsequent radiative decay to the repulsive $0^+_g$ $(^1\Sigma^+_g)$ ground state 
\begin{equation}
\label{eq:excdecay}
\mbox{Xe}_2^* \rightarrow 2 \mbox{Xe} +h\nu
\end{equation}
leads to the emission of the bands centered at 148 nm (first continuum) and 173 nm (second continuum).  
The former band is attributed to transitions from higher vibrational levels of the excimer level deriving from the A6s atomic configurations and is observed at very low pressure ($P<100$ Pa). The latter band, exploited in Xe gas based detectors, appears and dominates at higher pressure because it occurs after vibrational relaxation of the excited molecule.

Much less attention has been devoted to possible infrared (IR) or near--infrared (NIR) fluorescence, related to transitions between higher lying excited states of the rare-gas
dimers in pressurized gas, although some broad bands have been observed in noble gases at intermediate pressure ($0.05 <P<0.5$ MPa) \cite{kogo,arai69}, IR scintillation has been detected in liquefied Ar and Xe \cite{bressi00}, and atomic IR scintillation has been revealed in alpha--particle--excited noble gas  at low--pressure ($10<P<50$ kPa) \cite{solim88,lindblom88,lindblom88b}. 

Only recently, a strongly peaked NIR fluo\-re\-scen\-ce b\-and has been detected at moderately high pressure ($P<0.9$ MPa) in pure Xe and in an Ar--Xe mixture, both excited with an energetic electron beam \cite{Bor}. 
The band, centered around $\lambda^{-1}\approx 7800$ cm$^{-1},$ has been attributed to a transition between bound-- and dissociative levels of the Xe$_2$ excimer, which are lying higher in the energy relaxation pathway that eventually leads to the population of the excimer levels responsible of the VUV continua. 

The observed fluorescence spectra in pure Xe and in the Ar (90\%)--Xe (10\%) mixture are similar because they are both due to the de--excitation of the Xe$_2^*$ excimer. In fact, energy transfers from Ar to radiative states of Xe$^*$ occurs readily and heteronuclear three--body collisions of the type
\begin{equation}
\label{eq:heteronuclear}
\mbox{Xe}^* +\mbox{Xe} +\mbox{Ar} \rightarrow \mbox{Xe}_2^* +\mbox{Ar}
\end{equation}
lead to the formation of the Xe$_2$ species even in the mixture \cite{exc}.

The main feature shown by the observed NIR band is that its center is red--shifted upon increasing the gas pressure. The wave number of the band peak decreases linearly with the gas density $N$ up to $N\approx 2\times 10^{26} $ m$^{-3} = 7.5 N_{i.g.}.$ $N_{i.g.}=2.65\times 10^{25}$ m$^{-3}$ is the density of the ideal gas at temperature $T=273.15$ K and pressure $P=0.1$ MPa. 

The density dependence of the properties of the NIR band have been interpreted in terms of the influence of the environment on the excimer \cite{Bor}. It is assumed that the higher lying Rydberg--like levels of the excimer are modified by two density--dependent effects. The first one is the solvation effect due to the presence of the atoms of the host gas. Many of them are encompassed within the large orbit of the electron in the Rydberg--like state and screen the Coulomb interaction between the electron and the Xe$^{+}_{2}$ core, leading to a reduction of the difference between the energy levels. Thus, the solvation always moves the transition to longer wavelengths. 

The second effect is due to the quantum nature of the electron and to its interaction with the atoms of the host gas. On a large orbit, the electron is quite delocalized and its wave function spans a region containing several atoms, with which it is thus interacting simultaneously. This fact leads to a density--dependent shift of the energy of the electron ground state that is negative or positive depending on whether the electron-atom interaction is either attractive or repulsive \cite{fermi34}. 

This interpretative model of the observed red--shift gives good agreement with the experimental data in the investigated density range. 

It is to note that this phenomenon is related to the modifications of the molecular properties of Wannier--M\-ott--ty\-pe impurity states in high--density liquids \cite{raz70}, w\-hich have a unique parentage to the Rydberg states of an isolated molecule in a low density gas \cite{messing77,messing77b}. Moreover, it would be interesting to ascertain why NIR excimer fluorescence has not been detected yet in liquefied rare gases. 

In the previous measurements \cite{Bor}, it has been also observed that the width of the fluorescence band increases linearly with the gas density in both the pure gas and the mixture. It is known that a linear increase of the width of a band or line with increasing density is due to an increase of the collision rate, which decreases the mean lifetime of the atomic or molecular species. 
The analysis of the broadening of the band may therefore lead to the evaluation of the collision cross section of the excimer and to an estimate of its size that would allow the verification of the conditions, under which the model of the excimer as a Rydberg--like atom is valid. This analysis was not carried out in the previous mesurements \cite{Bor}.

For these reasons, we have carried out further and accurate measurements of the NIR fluorescence in the Ar--Xe mixture by extending the investigated density range up to $N\approx 6 \times 10^{26}$ m$^{-3}\approx 24 N_{i.g.},$  approximately 3 times larger than in the previous measurements. The goal is to confirm the validity of the interpretative model. 

\section{Experimental details}\label{sec:1}
The experimental goal is the detection and analysis of the NIR spectrum emitted by a gas
sample at high pressure and excited by an ionizing electron beam. We have used the same apparatus described in detail
in the previous paper \cite{Bor}. We report here only its
main features and some minor modifications. The electron beam is produced by a home made electron gun described in a
previous work \cite{belogurov00}. Electrons are injected in $50-$ns--long pulses with energy $\approx 70$ keV. Each pulse contains $\approx 
5$ nC charge and are repeated at a rate of $\approx 100$ Hz. The beam enters the gas chamber
through a $25-$$\mu$m--thick Kapton film. Upon
crossing the film, electrons lose about $30\%$ of their initial
kinetic energy and the beam in the gas partly loses its
directionality. The range of energetic electrons in the gas varies approximately between 5 and 100 mm, depending on the gas pressure.

The gas sample is kept at room
temperature inside a cylindrical stainless--steel chamber, which
is evacuated down to $10^{-5}$ mbar before being filled
with the gas. The gas is passed through an Oxisorb
cartridge for purification purposes. The pressure is measured with a Sensotec STJE/1833-35 pressure gauge with an accuracy of $\approx 5$ kPa. The cell is leak proof up 
to $4$ MPa.

The density is computed from the pressure values by using the equation of state of Argon found in literature \cite{rabi}. We have also calculated the density of the mixture by assuming an ideal gas
mixture law and using the appropriate  equation of state
\cite{rabi}. Although at room temperature Xenon is relatively
close to its critical temperature, the differences in the densities computed in either way are less than $1\%$ even at the highest
pressure.

 The emitted light exits the
chamber through a quartz window and it enters a FTIR (Fourier
Transform In\-fra\-Red) Michelson--type interferometer (Equinox $55$,
Bruker Optics). The interference signal is detected by an InGaAs
photodiode (G$5832-05$, EG\&G) kept at room temperature. The output signal of the
photodiode is pre--amplified by using an active integrator with a
conversion factor of $0.25$ mV/C and an integration constant of
$400$ $\mu$s. The signal is then digitized by a 16--bit A/D converter and is stored in a
computer.

Because the electron source is pulsed and each
pulse is too short to obtain a complete interferogram, a
step--scan technique is used \cite{haseth}. The movable mirror scans the whole range
in equal steps, whose amplitude is controlled by a reference He--Ne laser, until a complete
interferogram is acquired. At each intermediate position of the mirror several
light pulses are recorded and averaged together in order to
improve the S/N (Signal-to-Noise) ratio. The interferogram is converted to spectrum
by commercial software.

 The interferometer has been carefully aligned with respect to the IR source in order to maximize
the signal detected by the photodiode and, at the same time, to
have good interference. A first calibration has been im\-ple\-men\-ted 
by using an IR solid state laser with $\lambda\approx
900$ nm. 

As instrumental function we use a $3$--term
Blackmann--Harris function in order to suppress the wiggles in the
wings of the peaks in the spectrum. Because the shape of the spectral line, measured
at finite resolution, is the convolution of the natural and
instrumental line shapes, the natural line shape can only be
observed if the width of the instrumental line shape is small
compared to natural line width \cite{allard82}.

Because the full width at half maximum of the fluorescence band is $\approx 800$ cm$^{-1},$ we have set the intrumental resolution to 100 cm$^{-1}.$ Such
resolution has been chosen for two reasons. It must be not too low
in order to avoid excessive instrumental broadening and not too
high in order to limit the acquisition time to reasonable values.
In fact, the electron beam degrades the Kapton film that
consequently releases impurities that quench the formation of the excimers and the fluorescence intensity might decrease too much during each experimental run.

\section{Experimental results} \label{sec:2}

We have measured IR time--integrated
emission spectra in $90\%$ Ar--$10\%$ Xe gaseous mixture up to
$2.4$ MPa at room temperature. The corresponding gas densities are comprised in the range $(0.13\leq N\leq 6)\times 10^{26}$ m$^{-3}$. We have extended by a factor 3 the density range explored in the previous experiment \cite{Bor}.

\begin{figure}[htbp]\centering
\resizebox{0.75\columnwidth}{!}{%
\includegraphics{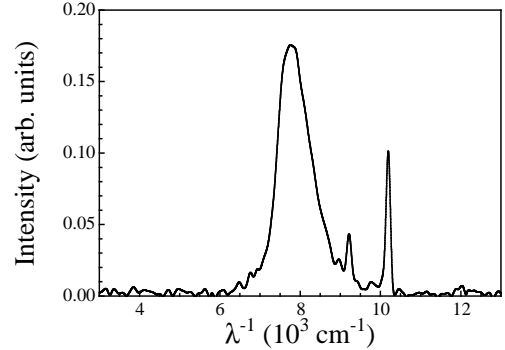}
}
\caption{\small Excimer  spectrum in Ar--Xe mixture at $T\approx298$ K and  $P =0.65$ MPa. }
\label{fig:1}
\end{figure}

In Figure \ref{fig:1} we show a typical spectrum obtained for $P= 0.65$ MPa.
 The main feature is the
broad continuum band centered at $\approx 7900$ cm$^{-1},$ very similar to that previously observed \cite{Bor}. 

This band has been observed also in pure Xe \cite{Bor} and is thus ascribed to a bound--free transition of the Xe$_{2}^{*}$ excimers.

In addition to the excimer band, two atomic lines can be observed in the spectrum, which are used as a reference to check the accuracy of the determination of the wave number of the fluorescence band, which is located nearby. These line are Xe I
lines present in the NIST databases \cite{Nist3}. The first one occurs at $\lambda^{-1} =  9224.8$ cm$^{-1}$ and is associated with the $6$p$[1/2]_{1}\, (^{3}S_{1})\rightarrow 6 $s$[3/2]_{1}\, (^{3}P_{1}^{o})$ transition \cite{sadeghi77}.  The second one, at  $\lambda^{-1} = 10194.0$ cm$^{-1},$ is associated  with the $6$p$[1/2]_{1}\,(^{3}S_{1})\rightarrow6$s$[3/2]_{2}\, (^{3}P_{2}^{o})$ transition. This latter line appears to be enhanced in Ar--Xe mixture
with respect to pure Xenon at low pressures \cite{galy} and also at higher pressures \cite{Bor}.

\begin{figure}[htpb]\centering
\resizebox{0.75\columnwidth}{!}{%
\includegraphics{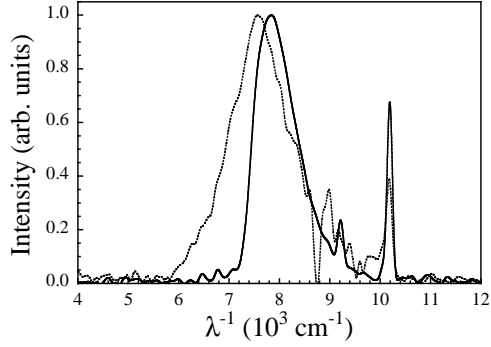}}
 \caption{\small Excimer spectra at two different pressures in the
Ar--Xe mixture. The spectra are normalized to unity at their maximum. Solid line: $P=0.15$ MPa. Dotted line: $P=2.4$ MPa. }
\label{fig:2}
\end{figure}
In Figure \ref{fig:2} we show two spectra recorded at two different pressures, namely $P=0.15$ MPa and $P=2.4$ MPa, which correspond to the densities $N= 0.38\times 10^{26}$ m$^{-3}$ and $N= 6.0\times 10^{26}$ m$^{-3},$ respectively. It is evident that the fluorescence peak undergoes a red--shift as the pressure is increased and that its width also increases. Such behavior confirms the results obtained previously for smaller densities \cite{Bor}.

\begin{figure}[htbp]\centering
\resizebox{0.75\columnwidth}{!}{%
\includegraphics{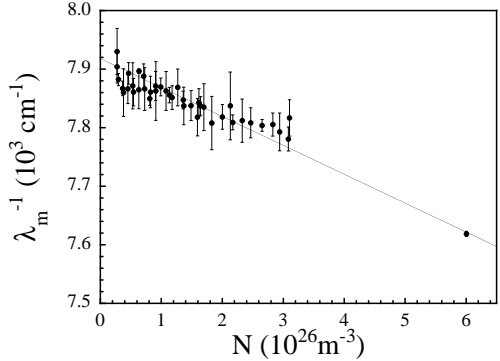}
} \caption{\small Red--shift of the center of the excimer band as a function
of $N.$  The error bar of the point at the highest density is of the same size of the dot.}
\label{fig:3}       
\end{figure}

The red--shift of the emission band can be best seen by plotting the position of its maximum as a function of the density, as shown in Figure \ref{fig:3}. In order to give an estimate of the statistical accuracy of the central wavenumber reported in this Figure, several lorentzian fits of the spectrum have been carried out on different data ranges around the peak. 
The central wavenumber of the band decreases linearly with $N.$ The present data are compatible, within the experimental accuracy, with the data of the previous work, in which a third--order polynomial fit of the spectrum around the maximum was used to determine its position \cite{Bor}.

\begin{figure}[htbp] \centering
\resizebox{0.75\columnwidth}{!}{%
\includegraphics{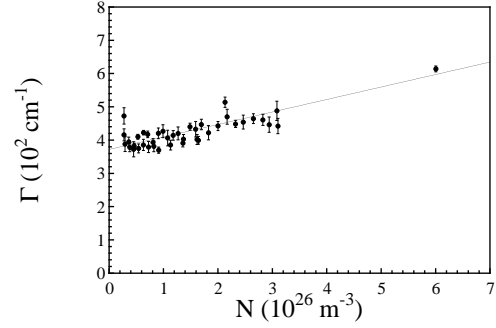}
}
 \caption{\small The excimer band half--width as a function of $N.$ }
\label{fig:4}
\end{figure}

Another important feature of the excimer band is its pressure broadening. In Figure \ref{fig:4}, $\Gamma,$ the half width at half maximum, is shown to increase linearly with $N.$ 
The error bars have been determined with a procedure similar to that used for Figure \ref{fig:3}.

\section{Discussion} \label{sec:3}

The excimer band maximum is centered at $\lambda_{m}^{-1}\approx
7900$ cm$^{-1}$ that corresponds to $\lambda_{m}\approx 1270$ nm
and to a transition energy of $E_{m}\approx 1$ eV. Such a large energy
value corresponds to electronic transitions. 

The large width of the NIR continuum suggests that the transition takes place between a
high--lying bound excimer level to a lower dissociative one. If the transition were involving two bound excimer levels, its width should be much sharper \cite{mulliken70}.

Moreover, the transition from the first excited excimer level to the ground
state corresponds to an energy of $\approx 8$ eV. Thus, the observed
emission does not involve the excimer ground state and is a transition between higher lying excimer levels, which are populated in one of the possible route, in which the initial electron kinetic energy is dissipated in the system. 

The observed wavelength range may correspond to the energy
difference between the stable excimer level ($0_{u}^{\pm} , $ $1_{u},$
$2_{u}$) of A$7$d$\pi$ configuration, correlated with the
Xe ($^{1}$S$_{0}$)\ $  $+Xe$^{*}$ ($5$p$^{5}[^{2}$P$_{3/2}]6$p) (or
Xe$^{*}$ ($5$p$^{5}$[$^{2}$P$_{3/2}$]$5$d)) dissociation limit,
and the energy level of the
Xe$(^{1}$S$_{0})$ + Xe$^{*}$ $(5$p$^{5}$ [$^{2}$P$_{3/2}$]\ $6$s). The
bound potential curves derive from $6$p and $5$d atomic levels,
which are the first excited states of Xe. The dissociative
potential curves derive from the $6$s atomic level \cite{mulliken70}.

It is easy to see from Figure \ref{fig:3} that the wavenumber $\lambda_{m}^{-1}$ of the center of the NIR continuum depends linearly on $N.$ The negative slope of the
linear fit clearly indicates red--shift. The parameters of the linear fit are
\begin{eqnarray}
\lambda^{-1}_{m,0} & = & (7918\pm 12)\> \mbox{cm}^{-1} \nonumber \\
d\lambda_{m}^{-1}/dN& = & -(4.94\pm 0.53)\times 10^{-23}\> \mbox{m}^{2}\nonumber
\end{eqnarray}
The previous experiment \cite{Bor}, carried out in a more restricted density range, 
yielded the values $(7837\pm 45)$ cm$^{-1}$ for the intercept at zero density, and $-(6.7\pm 3.3)\times 10^{-23} \mbox{m}^2$ for the slope, respectively. The present measurements are compatible, within the experimental accuracy, with the previous ones. However, the present data have a larger statistical significance owing to the higher number of experimental points and to the larger density range investigated.

We have developed a simple model that takes into account the interaction of the outer electron in the excimer with the atoms of the host gas in order to explain the observed density dependent red--shift \cite{Bor}. This model is also able to explain the reason why the red--shift of the excimer band in pure Xe ($d\lambda_{m}^{-1}/dN\approx -22 \times 10^{-23}$) is much larger than in the mixture.

The electronic structure of homonuclear excimers can be accurately described by an ionic
molecular core (Xe$_{2}^{+}$) plus an electron in a diffuse
Rydberg orbital, whose radius is much larger than the internuclear
distance \cite{kogo}. The evolution of a Rydberg state as a
function of density bears some similarities with the Wannier impurity states in
liquids or solids \cite{messing77b}. The Wannier excitons are related to
high--extravalence excitations of atomic and molecular impurities
\cite{messing77,messing77b}, which show a density--dependent red--shift when the
impurity is embedded in a high polarizable medium, while they are
blue--shifted when the host gas is less polarizable \cite{messing77}.

This analogy suggest to describe excimer decays in dense noble gases in the
same way as Wannier--Mott excitons are treated in liquids and solids \cite{raz70}.
The energy eigenvalues $E_{n}$ of the n$th$ Wannier impurity state
are given by \cite{raz70}
\begin{equation}
\label{eqn:eq1}E_{n}=I_{g}+P_{+}+V_{0}(N)-\frac{G_{0}}{K^{2}(N)n^{2}}
\hspace{0.8 cm} n=1, 2,\ldots
\end{equation}
where $I_{g}$ is the gas phase ionization potential,
$P_{+}$ is the polarization energy of the medium due to the ionic
positive core, and $N$ is the density of the host. $V_{0}(N)$ is
the energy of the bottom of the conduction band relative to the vacuum level.
 The last two terms of the energy eigenvalues in Equation \ref{eqn:eq1} turn out
to be density dependent.

The last term in Equation \ref{eqn:eq1} expresses
the coulombic interaction of the electron with the host atoms.
$G_{0}=13.6\,(m^{*}/m)$ (eV) is the exciton effective Rydberg
constant, which takes into account the effective electron mass
$m^{*}$ in the conduction band. $m$ is the free electron mass. $K(N)$ is the dielectric constant
of the medium that can be calculated with the
Lorentz--Lorenz formula $(K-1)/(K +2)=N\alpha/3\epsilon_{0}$,
where $\alpha$ is the atomic polarizability and
$\epsilon_{0}$ the vacuum permittivity.

The term $V_{0}(N)$ can be approximated at the present densities by the Fermi
potential that describes a quantum effect related to the large ratio of the wavelength
of the Rydberg electron to the interatomic distance. The interaction of the Ryd\-berg electron with the atoms
of the host gas produces a density--dependent shift of the
electron energy.
For large electron orbits it converges to a limit
that is proportional to the density \cite{fermi34}
\begin{equation}
\label{eqn:eq2}V_{0}(N)=\frac{2\pi \hbar^{2}}{m}Na
\end{equation}
where $\hbar = h / 2 \pi $ and $h$ is the Planck constant. $a$ is the scattering length for electron--atom interaction, which is negative for an attractive potential and positive for a repulsive one \cite{zecca96}. If $a<0$, $V_{0}(N)$ is
also negative. This implies a density--dependent red--shift of the spectral lines.

The Fermi energy shift $V_0 (N),$ probably, does not con\-tri\-bu\-te to the energy of the final state because, when the excimer decays to a lower--lying repulsive level, its radius in the final state is much smaller than that of the starting state and  its mean lifetime is too short for the outer electron to sample the environment. Therefore, the energy of the dissociative state is
written by dropping $V_{0}(N)$ in Equation \ref{eqn:eq1}
\begin{equation}
\label{eqn:eq3}E_{n}=I_{g}+P_{+}-\frac{G_{0}}{K^{2}(N)n^{2}}
\end{equation}
Furthermore, it is assumed that $G_{0}$ is the same in
the initial and final states. 

The central wavenumber of the
transition between the upper level of principal quantum number
$n_{i}$ and the lower dissociative level of principal quantum
number $n_{f}$ is
\begin{equation}
\label{eqn:eq4}\lambda_{m}^{-1}=\frac{G_{0}e}{hcK^{2}(N)}\left( \frac{1}{n_{f}^{2}}-\frac{1}{n_{i}^{2}}\right) + \frac{\hbar}{mc}Na
\end{equation}
$e$ and $c$ are the usual universal constants. At the densities of this experiment, the dielectric
constant can be approximated to first order in the density by
$K=1+N\alpha/\epsilon_{0}.$ 

By expanding $1/K^2$ to first order in $N$ and rearranging, we obtain
\begin{equation}
\label{eqn:eq5}\lambda_{m}^{-1}=\lambda_{m,0}^{-1}-\Big(\lambda_{m,0}^{-1}\frac{2\alpha}{\epsilon_{0}}-\frac{\hbar}{mc}a\Big)N
\end{equation}
where $\lambda_{m,0}^{-1}=G_{0}e/hc(n_{f}^{-2}-n_{i}^{-2})>0.$

The first term in
parentheses in Equation \ref{eqn:eq5} is a solvation energy and, thus, it always shifts the spectrum to longer wavelengths. The second term increases this shift if $a<0$ or decreases it if
$a>0$. In Equation \ref{eqn:eq5}, $\lambda^{-1}_{m,0}$ is a fitting parameter that
corresponds to the experimental value of the
wavenumber in the zero--density limit, i. e., the mean energy of
the transition.

We note that the concentration of the mix\-ture is large enough to
permit excimer formation, but also small e\-nough that the
Xe$_{2}^{*}$ excimers are surrounded by Argon atoms. Thus, in
Equation \ref{eqn:eq5}, on average, the atomic polarizability and
the scattering length of Ar are used. Their values are 
$\alpha=1.827\times 10^{-40}$ Fm \cite{maitland81} and
a$=-0.86\,\textrm{\AA}$ \cite{zecca96}, respectively. The theoretical slope is
thus $-(2\lambda^{-1}_{m,0}\alpha/\epsilon_{0}-\hbar a/mc)=-6.56\times 10^{-23}$
m$^{2},$ to be compared with the present value $-(4.94\pm 0.53)\times 10^{-23} \mbox{m}^2.$ The difference between the experimental and the
theoretical determinations of the slope might be traced back to
the approximation of neglecting the Fermi contribution $V_{0}(N)$
in the final state.

Another important piece of information that can be obtained from
the experimental data is the density dependence of the width of the excimer
band. The increase of collision frequency by increasing
pressure causes a broadening of the band. By inspecting Figure \ref{fig:4}, we observe that $\Gamma$ is well fitted to the straight line
\begin{equation}
\label{eqn:eq6}\Gamma=\Gamma_{0}+\gamma N
\end{equation}
where $\Gamma_{0}=(3.72\pm 0.05)\times 10^{2}$ cm$^{-1}$ is the
half--width in the zero-density limit, and $\gamma=(0.37\pm
0.05)\times 10^{-22} $ m$^{2}$ (or $\gamma=(0.37\pm0.05)\times
10^{-24}$ cm$^{-1}$m$^{3}$ in practical units) is positive and clearly indicates broadening.

 The data reported in
Figure \ref{fig:4} have been corrected for the instrumental line
width $\Gamma_{ins}\approx 100$ cm$^{-1}$. The observed width
$\Gamma_{exp}$ is related to the true width $\Gamma$ and the
instrumental one $\Gamma_{ins}$ as
\begin{equation}
\label{eqn:eq20}\Gamma=\sqrt{\Gamma_{exp}^{2}-\Gamma_{ins}^{2}}
\end{equation}
$\Gamma_{ins}<\Gamma_{exp}$ and the influence of $\Gamma_{ins}$ on the 
 true width is $3\%$ at most.
 
 In the case of an atomic line, the width
represents the total number of spontaneous transition per second
from an initial excited energy state to a final one and it
 is related to the mean lifetime $\tau$  of the state by the
uncertainty principle. For a continuous band, however, the
transition involves a broad continuum of final energy states, so
that the width is an estimate of their energy range. In any case,
there is an evident pressure dependent band broa\-de\-ning
similar to line broa\-de\-ning in pure atomic transition \cite{margenau36}.

In addition to instrumental broadening, Doppler-- and collision broadening, which will be described in Appendix,
may affect the width of the band. 
In the present case, the Doppler effect gives a broadening $\Gamma_D\,\approx 
1.7\,\times 10^{-2}$ cm $^{-1}$ and is completely negligible.  Impact broadening is proportional to the density with a proportionality coefficient $\gamma^\prime \approx 0.2\times 10^{-24}$ m$^2,$ two orders of magnitude smaller than the observed $\gamma $ and is therefore neglible, too.

At the densities of this experiment, the average interatomic distance is comparable to the atomic 
van derWaals radii and the results of the impact broadening theory cannot be applied.

 In this situation, we characterize the broadening according to pure kinetic theory by introducing
an effective total scattering cross--section that can be determined
from the experimental band width.
This procedure will enable us to give an estimate of the size of the excited state of Xe$_{2}^{*}$
investigated in this experiment. 

The measured half--width $\Gamma$
is related to the mean lifetime $\tau$ of the species from the
relation
\begin{equation}
\label{eqn:eq10}2\pi
c\Gamma=\frac{1}{\tau}=\frac{1}{\tau_{0}}+\frac{1}{\tau_{c}}
\end{equation}
where $1/\tau_{0}$ is the inverse mean lifetime in the
zero--density limit and $1/\tau_{c}=\sqrt{2}N\sigma\overline{v}$
is the collision frequency. $\overline{v}=\sqrt{8RT/\pi M}$ is the
average velocity. $T$ is the temperature, $R$ is the gas constant, and $M$ is the reduced molecular mass of the 
Xe$_2^*-$Ar system \cite{maitland81}. $\sigma$ is the total cross--section and can
be determined from the slope of the linear fit of $\Gamma$ as a function of $N.$

 The fit of the experimental data yields $\sigma = 1.15\times 10^{-16}$ m$^{2}$.
 The cross section is related to the  collision diameter $\rho,$ which represents the
distance between the particle centers at the instant of collision, by $\sigma=4\pi\rho^{2}$, as it is
derived from quantum mechanics in the limit of the low--energy hard--sphere scattering \cite{schiff84}.

The collision
diameter for the Xe$_{2}^{*}$ excimer--Ar atom system turns out to be $\rho=3.02$ nm.
The Argon hard sphere diameter is $d_{Ar}=0.33$ nm
\cite{maitland81}. Thus, the excimer diameter turns out  to be $d_{exc}=5.71$ nm.
Such a large value supports the assumption that the orbit of the
outermost electron is large enough to encompass several atoms of
the host gas, thus justifying the approximation of continuum dielectric screening used in the model.

\section{Conclusions} \label{sec:4}

We have carried out accurate measurements of the NIR fluorescence band of Xe$_{2}^{*}$ excimers produced in an e\-lec\-tron--be\-am--ex\-ci\-ted Xe--Ar gas\-eous mixture as a function of the gas density, up to a maximum density $N_m$ $ \approx 24 N_{i.g.},$ where $N_{i.g}$ is the density of the ideal gas at standard temperature and pressure. We have thereby extended by a factor 3 the density range explored in the previous measurements \cite{Bor}, which spanned the density up to 7.5 $N_{i.g.}.$

We have observed a red--shift of the excimer spectra that is linear with $N$ up to $N_m.$
In the density range common to both series of measurements, the results are compatible with each other within the experimental accuracy. Therefore, the  present measurements essentially confirm and extend the results obtained in the previous experiment \cite{Bor}.

 We have quantitatively 
explained the density dependent shift of the maximum with a model that puts into evidence the interaction between the outermost electron 
of the
Xe$_{2}$ molecule and the atoms of buffer gas. A great polarizability and a
negative scattering len\-gth of the buffer gas produce a red--shift. This fact is well confirmed by experimental measurements in pure Xe gas \cite{Bor}, for which these requirements are met. 

The density dependent red--shift of the excimer fluorescence in this system, as far as its dependence on the electron--atom scattering length and on the atomic polarizability of the gas atoms, bears strong analogies with the shift of the high--extravalence 
excitation transition to the $7$s$[3/2]_{1}$ state of Xe impurity in fluid Ar and 
Ne \cite{messing77}.

 Finally, for the first time the analysis of the 
 broadening of the Xe$_{2}^{*}$ excimer band has been carried out. 
 Collisions are the most important causes of the band broadening. From the density dependence of the band width the collision cross section of the Xe$_2^* -$Ar system can be computed and 
the molecular radius can be determined. Its value for the Xe$_{2}^{*}$ molecule turns out to be 2.85 nm. Such a large value lends credence to the model of the excimer as an ionic core plus an electron in a Rydberg--like state, whose very large orbit encompasses many atoms of the host gas.

\section*{Appendix}\label{appendice}
In this Appendix we briefly describe the main mechanisms that can concur to the broadening of a line or band.
The first one is the Doppler effect, which is independent of the density of the gas. 
It causes the emission band to take on a gaussian shape, whose width is \cite{margenau36}
\begin{equation}
\label{eqn:eq21}\Gamma_{D}=2\lambda_{m}^{-1}\sqrt{\frac{2\ln 
2RT}{Mc^{2}}}
\end{equation}
$c$ is the 
speed of light and $M$ is the reduced molecular mass of the
Xe$_{2}^{*}-$Ar system.
 For this experiment, $T\approx 300$ K, $\lambda_m^{-1}	
\approx 7900 $ cm$^{-1}$ and  
$\Gamma_D \approx 1.7\times 10^{-2} \mbox{cm}^{-1}$ is completely negligible with respect to the band full--width.

The other effect is collision broadening. The {\em impact theory} has been developed in the limit of low densities \cite{allard82,margenau36,vleck45,lewis59}, at which atoms or molecules interact via van der Waals forces.
The potential due to a long--range attraction, as firstly derived by
London \cite{london30}, is
\begin{equation}
\label{eqn:eq7}
W(r)=-C_{6}/r^{6}
\end{equation}
where $r$ is the distance
between the two interacting particles $a$ e $b.$
$C_6$ is given by \cite{maitland81}
\begin{equation}
\label{eqn:c6}
C_{6}=
- \frac{3\hbar}{\pi\left(4\pi\epsilon_0\right)^2}
\int\limits_{0}^\infty 
\alpha_a\left( i \omega\right) \alpha_b \left( i \omega\right) \mathrm{d}\omega
\end{equation}
where $\alpha_{a,b}(i\omega)$ are the atomic
polarizabilities of the two different atoms at the imaginary
frequency $i\omega$.
 The expected broadening depends on $C_{6}$ and linearly on the density \cite{allard82}
\begin{equation}
\label{eqn:eq9}\Gamma_{vdW}=\frac{\beta}{2\pi
c}\overline{v}^{3/5}\Big(\frac{C_{6}}{\hbar}\Big)^{2/5}N \equiv\gamma^{'}N
\end{equation}
where $\beta=(3\pi/8)^{2/5}(\pi^{3/2}/4)\cos(\pi/5)\simeq 1.2$ is
a numerical factor \cite{allard82}. 

It is possible to calculate $\gamma^{'}$ if $C_{6}$ is known and
compare it with the $\gamma$ value of the experimental fit in
order to estimate the influence of van der Waals broadening.

 For the Xe$_{2}$--Ar system $C_{6}$ is unknown. However, estimates 
of $C_{6}$ for the system Xe--Ar are available from density
functional theory, $C_{6}=121\times 10^{-79}$ Jm$^{6}$
\cite{and96}. We thus obtain $\gamma^{'}=0.2\times 10^{-24}$
m$^{2}\ll\gamma=0.37\times 10^{-22}$ m$^{2}$.
The result is nearly two orders of magnitude smaller than the
experimental $\gamma$ and the van der Waals broadening is negligible in our experiment.

It is to remark the fact that we have used a $C_{6}$ value that is not
the correct one. There are indications in literature \cite{and96}
that $C_{6}$ for a system consisting of a noble gas plus a
diatomic molecule is approximatively twice as large as $C_{6}$ for
the system, in which the molecule is replaced by its parent atomic
species. Even using such a larger
$C_{6}$ value this correction remains negligible.

We have considered here only dipole-dipole interaction. In
order to give a more complete treatment of impact broadening,
repulsive interaction and higher--order multipole interactions should be taken into 
account \cite{hind66,hind70} but they give even more
negligible contributions.

 Finally, this theory is valid
only for atoms in the wings of the potential, for which the
atom--atom distance is grea\-ter than the van der Waals radius,
namely, at pressures much lower than those of this experiment, in which
the average interatomic distance is comparable with the van der
Waals radius.

\bibliography{ArXe}

\end{document}